\documentclass[a4paper,11pt]{article}
\pdfoutput=1 

\usepackage{jinstpub} 

\usepackage{amsmath}

\usepackage{graphicx}  

\title{ \boldmath The Mu2e crystal calorimeter}

\author[d]{ N. Atanov}
\author[d]{ J. Budagov }
\author[c] { F. Cervelli }
\author[a]{ F. Colao}
\author[a]{ M. Cordelli}
\author[a]{ G. Corradi}
\author[a]{ E. Dan\`e}
\author[d]{ Y. Davidov}
\author[c]{ S. Di Falco}
\author[a]{ E. Diociaiuti}  
\author[b,c]{ S. Donati}
\author[a]{ R. Donghia}
\author[e]{ B. Echenard}
\author[a]{ S. Giovannella}
\author[d]{ V. Glagolev}
\author[f]{ F. Grancagnolo}
 \author[a,1]{ F. Happacher\note{Corresponding author.}}
  \author[e]{ D. Hitlin}
  \author[a,h]{ M.Martini}
\author[a]{ S. Miscetti}
\author[e]{  T. Miyashita}
\author[c]{ L. Morescalchi}
\author[g]{ P. Murat} 
\author[c]{ E. Pedreschi}
\author[c]{G.~Pezzullo}
\author[e]{ F. Porter}
\author[a]{ A. Saputi}
\author[a,h]{ I. Sarra}
\author[c]{F. Spinella}
\author[f]{ G. Tassielli}
 
  \affiliation[a]{INFN Laboratori Nazionali di Frascati , via Enrico Femri 40, Frascati, Italy} 
 \affiliation [b]{ Departement of Physics, University of Pisa, Largo B. Pontecorvo 3, Pisa , Italy} 
 \affiliation[c]{ INFN sezione di Pisa, Italy, Largo B. Pontecorvo 3, Pisa , Italy}
 \affiliation[d]{ Joint Institute for Nuclear Research, Joliot-Curie 6, Dubna, Russia}
  \affiliation[e]{Departement of Physics, California Institute of Technology, 1200 E California Blvd, Pasadena (CA), USA}
\affiliation[f]{  INFN sezione di Lecce, Via Arnesano 73100, Lecce, Italy}
\affiliation[g]{  Fermi National Accelerator Laboratory, Main Entrance Rd, Batavia (IL), USA}
\affiliation[h]{  Department of Energy, University Guglielmo Marconi, via Plinio, 44, 00193 Roma, Italy}

\emailAdd{fabio.happacher@lnf.infn.it}

\abstract{The Mu2e Experiment at Fermilab will search for coherent, neutrino-less conversion of negative
muons into electrons in the field of an Aluminum nucleus, $\mu^- + Al \to e^- +Al$.  Data collection start is planned for the end of 2021. 

The dynamics of such charged lepton
flavour violating (CLFV) process is well
modelled by a two-body decay, resulting in a mono-energetic electron with 
an energy slightly below the muon rest mass.
If no events are observed in three years of running, Mu2e will set an upper limit on
the ratio between the conversion and the capture rates
R$_{\mu e} =  \frac{\mu^- + A(Z,N) \to e^- +A(Z,N)}{\mu^- + A(Z,N) \to \nu_{\mu} ^- +A(Z-1,N)} $ of $\leq
6\ \times\ 10^{-17}$ (@ 90$\%$ C.L.). 

This will improve the current
limit of four order of magnitudes with respect to the previous best experiment.

 Mu2e complements and extends the current search for $\mu \to e \gamma$ decay at MEG as well
as the direct searches for new physics at the LHC. 
The observation of such  CLFV process  could be clear evidence for New Physics beyond 
the Standard Model. Given its sensitivity, Mu2e will be able to probe New Physics at a scale
 inaccessible to direct searches at either present or planned high 
energy colliders. 

To search for the muon conversion process, a very intense pulsed beam of negative muons 
($\sim 10^{10} \mu/$ sec) is stopped on an Aluminum target inside a very long solenoid 
where the detector is also located. The Mu2e detector is composed of a straw tube tracker 
and a CsI crystals electromagnetic calorimeter.  An external veto for cosmic rays  surrounds 
the detector solenoid.  In 2016, Mu2e has passed  the final approval stage from DOE and 
has started its construction phase. 

An overview of the physics motivations for Mu2e, the current status of the experiment and the 
required performances and design details of the  calorimeter are presented. }


\arxivnumber{1234.56789} 

\proceeding{N$^{\text{th}}$ Workshop on X\\
  when\\
  where}

\begin{document}
\maketitle
\flushbottom

\section{Charged Lepton Flavor Violation and muon to electron conversion}
\label{sec:intro}

Within the Standard Model (SM), transitions in the lepton sector between charged and neutral particles preserve flavor,  since the neutrinos are considered massless.
Even considering the discovery of neutrino oscillations, in the minimal extension of SM, 
the predicted branching ratios of Charged Lepton Flavor Violation (CLFV) 
processes in the muon sector are smaller than 10$^{-50}$.

No CLFV process has been observed yet, so any experimental detection 
would be a clear signature of New Physics (NP) beyond the Standard Model.
One of the most promising process for probing CLFV is the coherent muon conversion 
in the field of a nucleus, $\mu$~N~$\rightarrow$~e~N. 
In this process the nucleus is left intact and the resulting electron has a monochromatic 
energy slightly below the muon rest mass ($\sim$~104.96 MeV for Al), due to the nucleus recoil. 

The Mu2e experiment \cite{tdr} is designed to improve the current limit on the conversion rate, R$_{\mu e}$, by 4 
orders of magnitude over the SINDRUM II experiment \cite{sindrum}. 
R$_{\mu e}$  is defined as the ratio between the number of electrons from the conversion process 
and the number of captured muons:
\begin{equation*}
R_{\mu e} = \frac{\mu^- \thinspace N(Z,A) \rightarrow e^- \thinspace N(Z,A)}{\mu^- \thinspace N(Z,A) \rightarrow\nu_{\mu} \thinspace N(Z-1,A)} 
\end{equation*}
where, in the Mu2e case, N(Z,A) is an Aluminum nucleus. 

Many NP scenarios, like SUSY, Leptoquarks, Heavy Neutrinos, GUT, Extra Dimensions or Little Higgs, 
predict significantly enhanced values for R$_{\mu e}$, allowing the detection of the process with the expected Mu2e sensitivity~\cite{marciano}.

\begin{figure}
\begin{center}
\includegraphics[width=0.8\columnwidth]{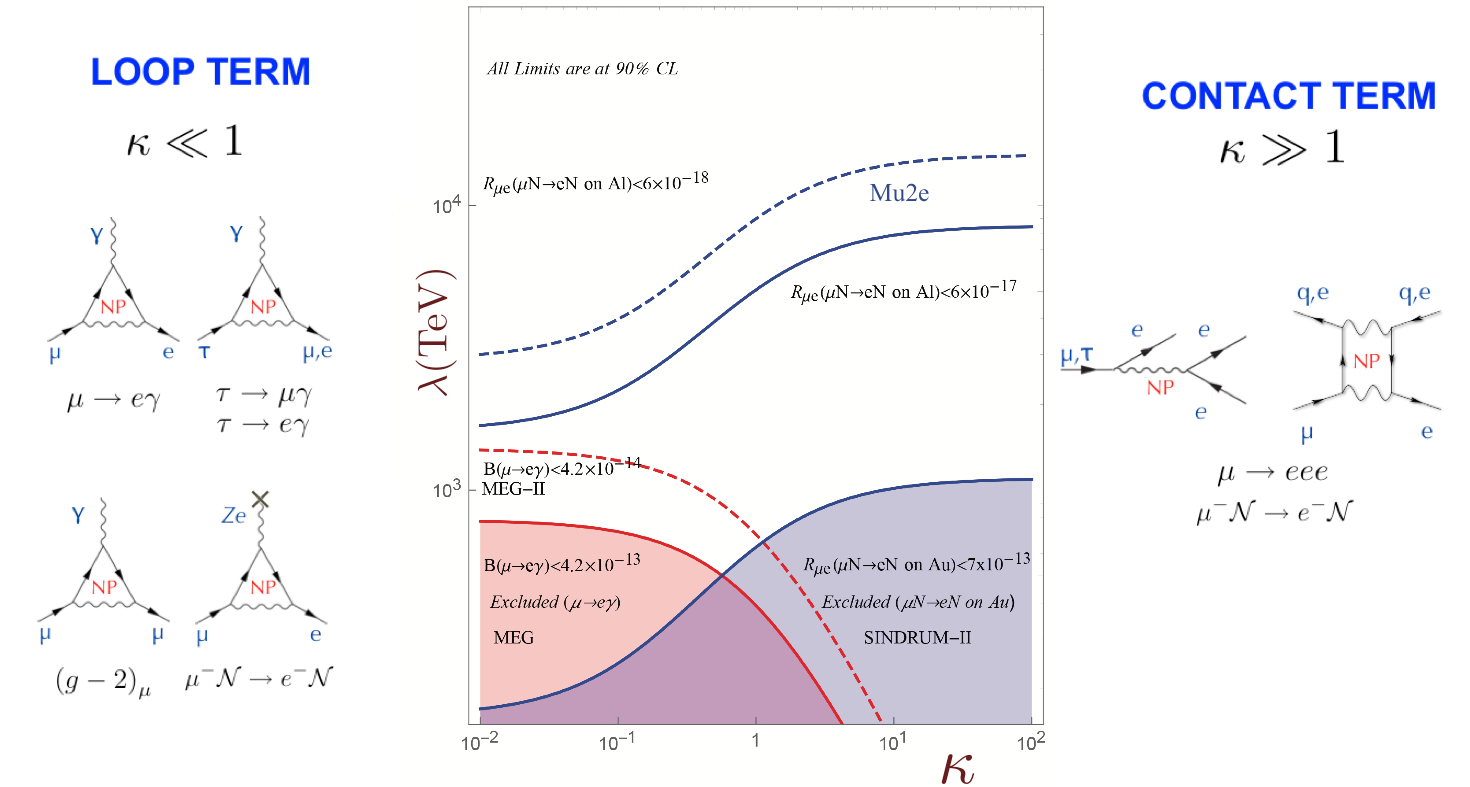}
\caption{\label{fig:degouvea} 
Sensitivity of $\mu \to e \gamma$, $\mu \to e$ transition and $\mu \to 3e$ to the 
scale of new physics $\Lambda$  as a function of the parameter $\kappa$ .
The shaded areas are excluded by present limits. On the left (right) side, the dipole (four-fermion) diagrams
are shown for the different processes.}  
\end{center}
\end{figure}

A model independent description of the CLFV transitions, for NP models, is provided by
an effective Lagrangian~\cite{deGouvea} where the different processes are divided in dipole amplitudes
and contact term operators. The $\mu \to e \gamma$ decay is mainly sensitive to the 
dipole amplitude, while $\mu \to e$ conversion and $\mu \to 3e$ receive contributions
also from the conctact interactions. It is possible to parametrise the interpolation between the
two amplitudes by means of two parameters~\cite{deGouvea}:
$\Lambda$, which sets the mass scale, and $\kappa$, which governs the ratio of the four fermion
to the dipole amplitude. 
For $\kappa <\!\!\!< 1 (>\!\!\!> 1)$ the dipole-type (contact) operator dominates.
Figure~\ref{fig:degouvea} summarises the power of different searches to explore this parameter space~\cite{bob}.

Present experimental limits already excluded lepton flavour violation up to a mass scale up of  $\Lambda < 700$~TeV. 
The interpretation of an eventual direct observation of NP at LHC 
will have to take into account precise measurements (or constraints) from MEG~\cite{meg} and Mu2e:
the comparison between these determinations will help pinning down the underlying theory.

\section{Muonic Aluminum atom}

When negative muons stop in the Aluminum target, they are captured in an atomic
excited state. They promptly fall to the ground state, then 39\% of them decay in
orbit, $\mu^- \to e^- \bar{\nu_e} \nu_{\mu}$  (DIO), while the remaining 61\% are captured on the nucleus. Low energy photons, neutrons and
protons  are emitted in the nuclear capture process and constitutes an
environmental background that produces a ionisation dose and a neutron
fluency on the detection systems
as well as an accidental occupancy for the reconstruction program.

The kinematic limit for the muon decay in vacuum is at about 54 MeV, 
but the nucleus recoil generates a long tail that has the 
endpoint exactly at the conversion electron energy.

\begin{figure}[hbt]
\begin{center}
\includegraphics[width=0.45\columnwidth]{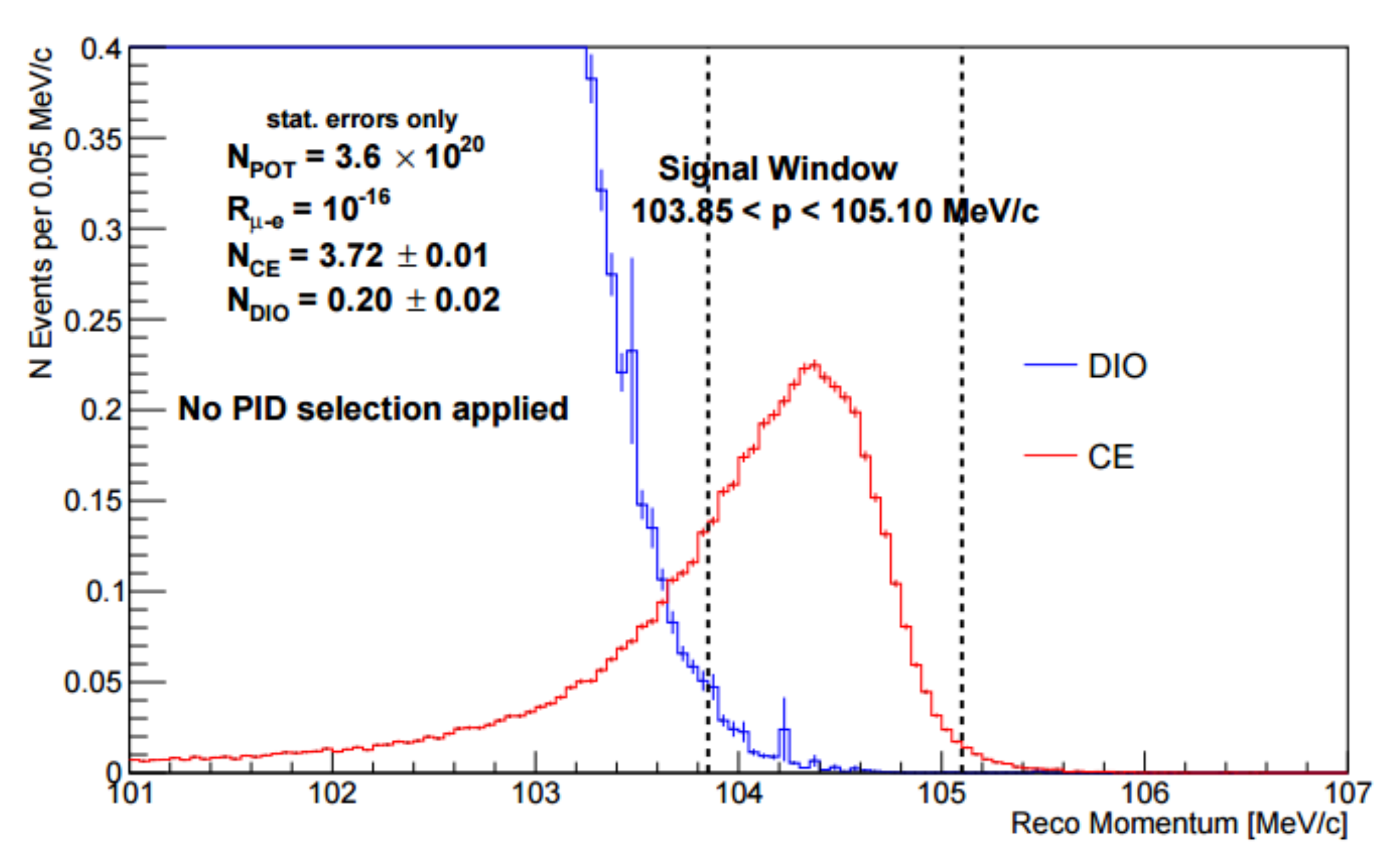}
  \caption{\label{fig:tail} Full simulation of DIO and CE events for an assumed $R_{\mu e}$ of 10$^{-16}$.} 
\end{center}
\end{figure}

DIO electrons are an irreducible background that have to
be distinguished by the mono-energetic conversion electron (CE). The finite tracking
resolution and the positive reconstruction tail has a large effect
on the falling spectrum of the DIO background that translates in a
residual contamination in the signal region as shown in Fig~\ref{fig:tail}.

\section{The Mu2e experimental apparatus}
\begin{figure}[ht]
\begin{center}
\includegraphics[width=1\columnwidth]{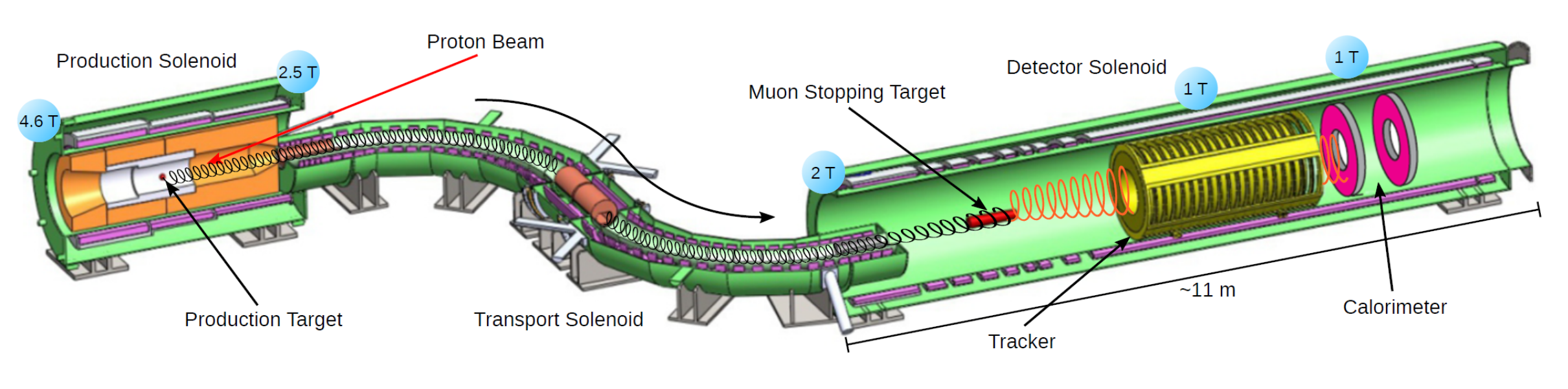}
\caption{\label{fig:mu2esetup} Layout of the Mu2e experiment.}
\end{center}
\end{figure}

The Mu2e apparatus consists of three superconductive solenoid magnets, 
as shown in Figure~\ref{fig:mu2esetup}: the Production Solenoid (PS), 
the Transport Solenoid (TS) and the Detector Solenoid (DS). 

The proton beam interacts in the PS with a tungsten target, producing mostly pions and muons. 
The gradient field in the PS increases from 2.5 to 4.6 T
in the same direction of the incoming beam and opposite to the
outgoing muon beam direction. This gradient field
works as  a magnetic lens to focus charged particles into the transport
channel. The focused beam is constituted by muons, pions with
a small contamination of protons and antiprotons. When the beam passes through the S-shaped TS,
 low momentum negative charged particles are selected and delivered to the Aluminum stopping 
targets in the DS. Electrons from the $\mu$-conversion (CE) in the stopping target are captured
 by the magnetic field in the DS and transported through the Straw Tube Tracker, 
 that reconstructs the CE trajectory and its momentum. The CE then strikes the Electromagnetic 
 Calorimeter, that provides independent measurements of the energy, the impact time and the position. 
 Both detectors operate in a 10$^{-4}$ Torr vacuum and in an uniform 1 T axial field. 

A Cosmic Ray Veto (CRV) system covers the entire 
DS and half of the TS, as shown in Figure~\ref{fig:calorimeter} (right).
  

Additional details on the Mu2e apparatus can be found in~\cite{tdr}.

\section{The Mu2e Tracker and Cosmic Ray Veto }
\begin{figure}[ht]
  \begin{center}
    \begin{tabular}{cc} \\
      \includegraphics[width=0.46\columnwidth]{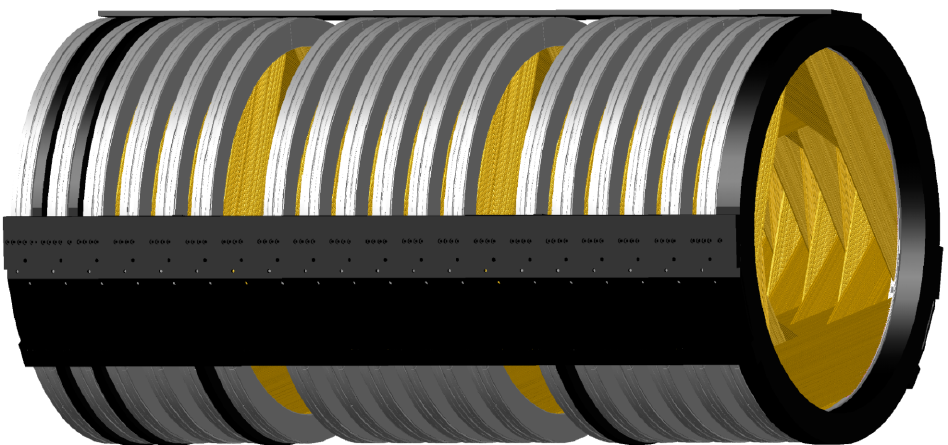}  &
      \includegraphics[width=0.50\columnwidth]{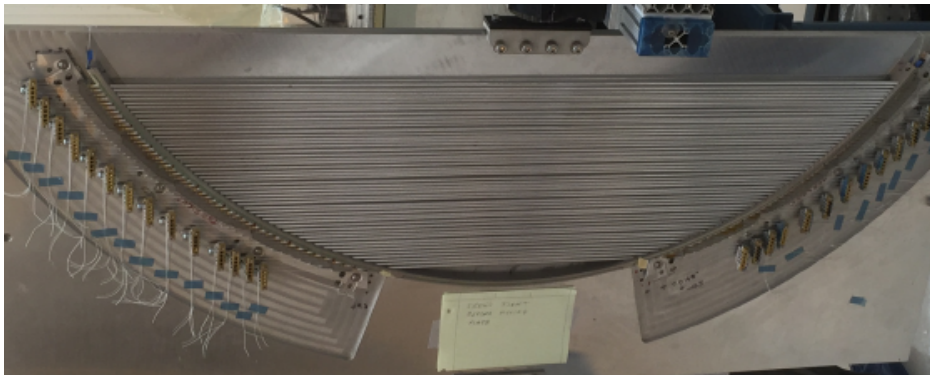} \\
    \end{tabular}
    \caption{\label{fig:tracker} (Left) Sketch of the Mu2e straw tracker system.
    (right) Picture of the first prototype built for straw tube panel. }
\end{center}
\end{figure}

The tracking detector is made of low
mass straw drift tubes oriented transversally to the
solenoid axis. The detector consists of about 21000 straw tubes arranged
in 18 stations, as shown in Figure~\ref{fig:tracker} (left). Each tube is of
5 mm in diameter and contains a 25~$\mu$m sense wire. The straw walls are
made  of Mylar and have a thickness of 15~$\mu$m. 

\begin{figure}[ht]
  \begin{center}
\includegraphics[width=0.40\columnwidth]{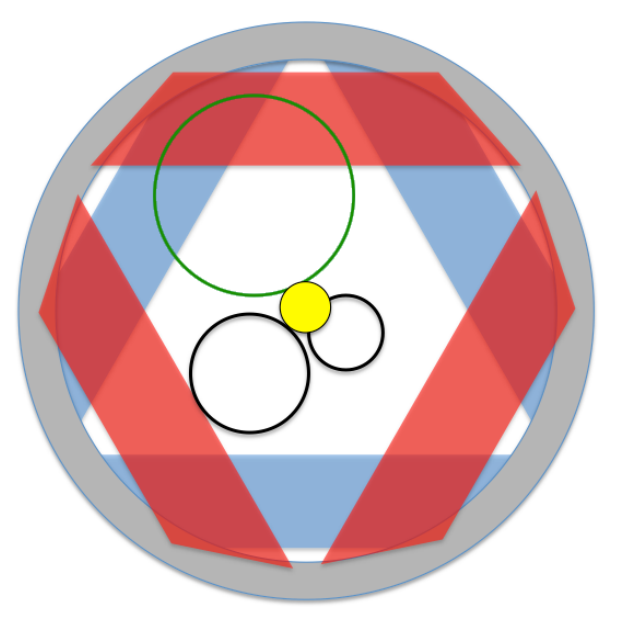} 

\caption{\label{fig:helics} Transverse view of the Tracker active area. Only tracks emerging from the stopping target (yellow spot) with 
momentum $p > 55$ MeV/c (green circles) leave hits in the straw tubes. Lower momentum tracks (black circles) leave the detectors undetected.}
\end{center}
\end{figure}
The gas used is a 80:20 mixture of Argon-CO$_2$. The tracker is around
3 m long and measures
the momenta of the charged particles from the reconstructed trajectories
using the hits detected in the straw. As shown in Figure~\ref{fig:helics}, a  circular inner un-instrumented region inside the tracker
makes it insensitive to charged particle with momenta below 55 MeV$/$c. 
Indeed its acceptance is optimised to identify $\sim~100$~MeV electrons.
Each straw tube is instrumented on both sides with TDCs to measure  the particle crossing time and ADCs to measure the specific energy loss $d$E/$d$X  used to separate electrons from highly ionizing particles. The   Momentum resolution for 105 MeV electrons is expected to be better than 180 keV$/$c, enough to suppress background electrons coming from the decays of muons captured by Al nuclei and from  DIO.

In Figure~\ref{fig:tracker}~(right),
an example of the first panel prototype built is shown. 

One major background source for Mu2e is related to cosmic ray muons faking CEs when
interacting with the detector materials. 
In order to reduce their contributions to below 0.1 event in the experiment lifetime, 
the CRV system is required to get a vetoing efficiency of at least 99.99\%
for cosmic ray tracks while withstanding an intense radiation environment. The basic element
of the CRV is constituted by four staggered layers of scintillation bars,
each having two embedded Wavelength Shifting Fibres readout by means of
(2$\times$2)~mm$^2$ SiPM.

\section{The Mu2e Calorimeter}

The electromagnetic calorimeter \cite{miscetti} is a high granularity  crystal-based calorimeter 
needed to:
\begin{itemize}
\item identify conversion electrons
\item provide particle identification to suppress muons and pions faking conversion electrons
\item add trigger capabilities
\item add seed positioning and timing in the track reconstructions
\end{itemize}

It is composed of two annuli  with inner and outer radii of  37.4 cm  and 66  cm respectively, filled by
pure CsI scintillating crystals and is placed downstream
the tracker. Each annulus is composed of 674  crystals of ($34\times34\times200$)~mm$^3$ dimensions,  each readout by two custom arrays of 2$\times3$ $6\times6$ mm$^2$ UV-extended Silicon Photomultiplier (SiPM); the SiPM are optimised to increase the quantum efficiency for 315 nm photons, the fast emission component of the scintillating process of CsI crystals.
The granularity and crystal dimensions have been optimised to maximise light collection for readout photosensor, time and energy resolutions and take under control particles pile-up.
Each crystal is wrapped with 150 $\mu$m Tyvek 4173D to maximise light collection.

\begin{figure}
  \begin{center}
    \begin{tabular}{ll} \\
      \includegraphics[width=0.45\columnwidth]{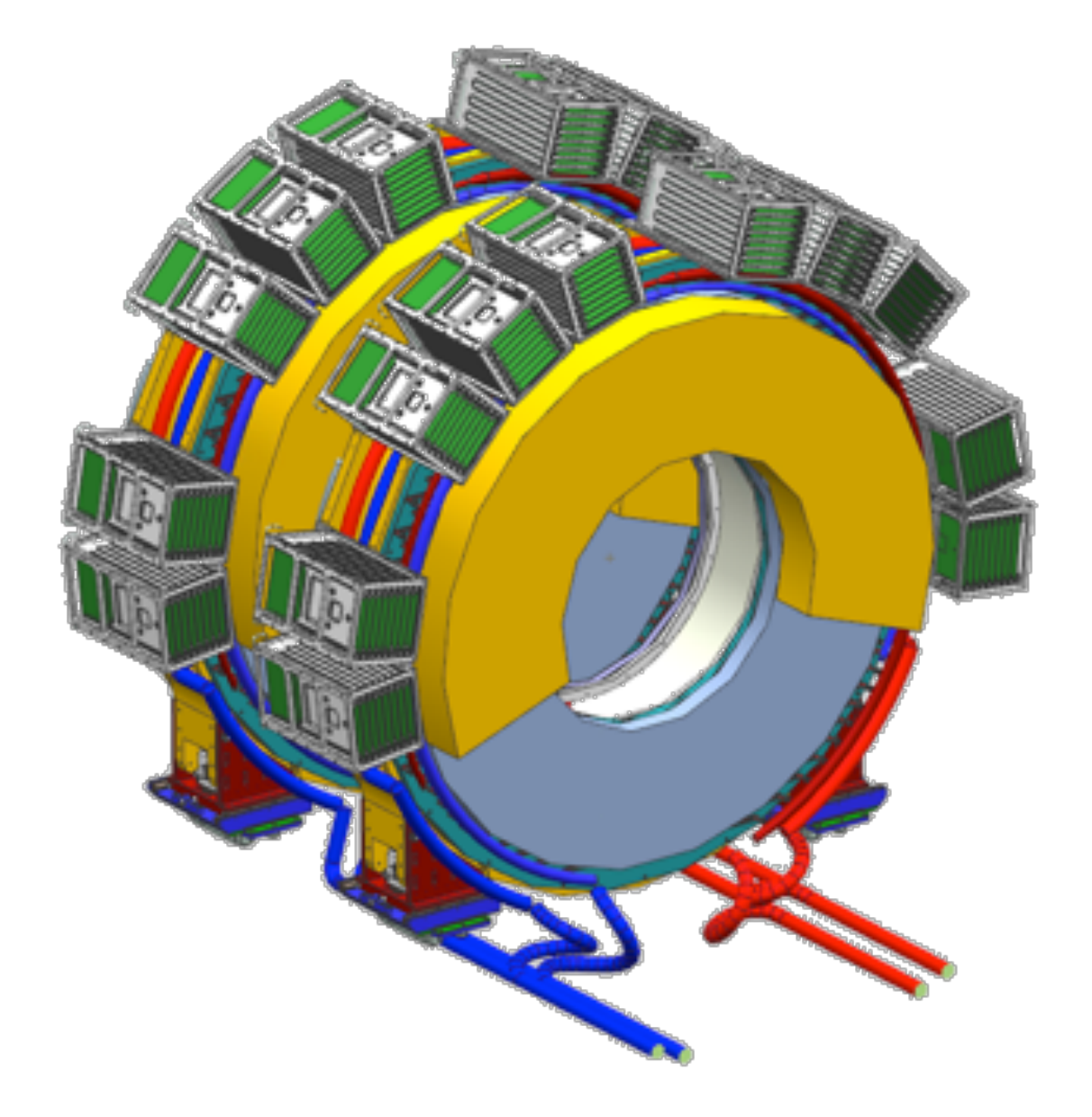}  &
      \includegraphics[width=0.5\columnwidth]{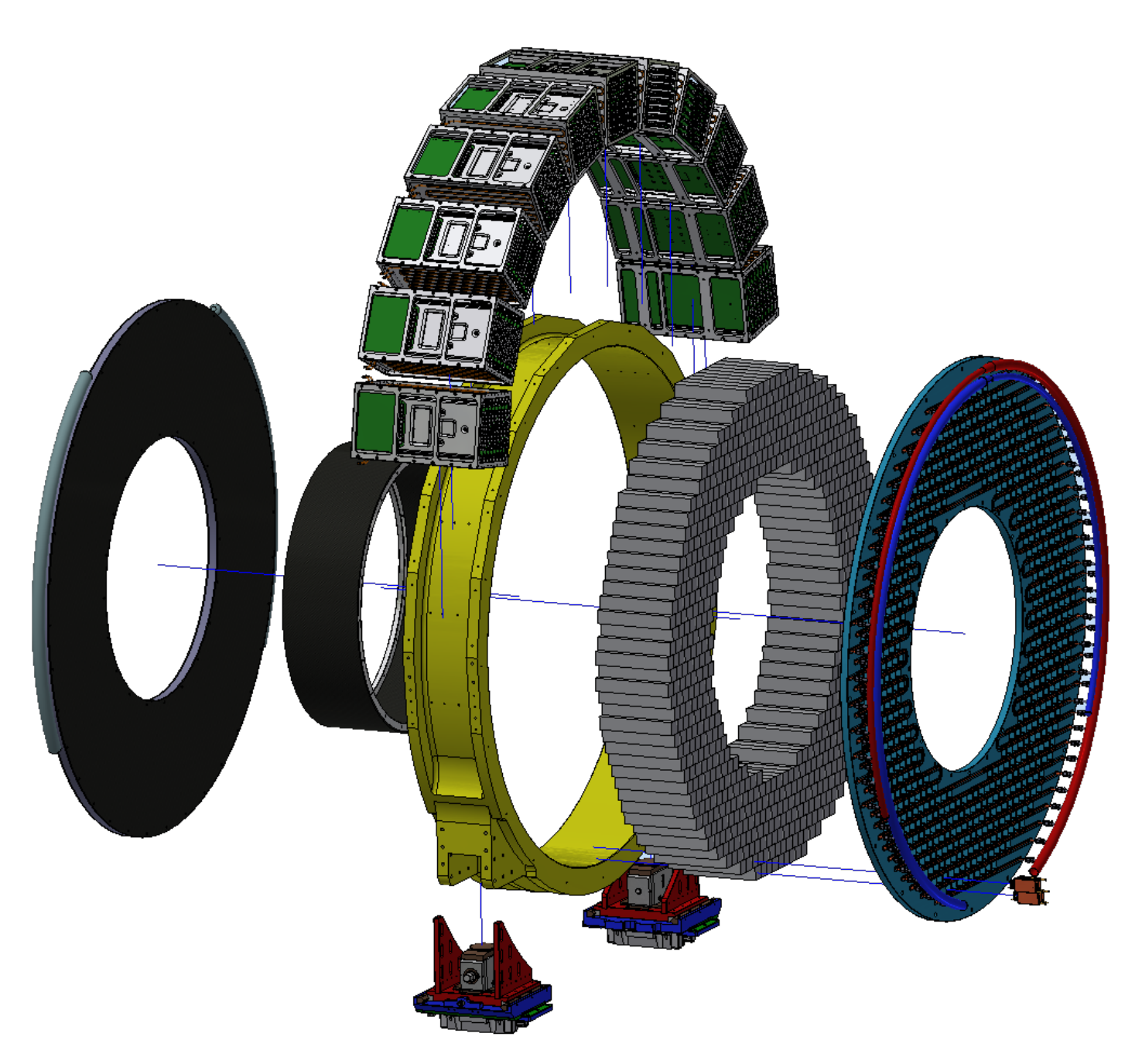} \\
    \end{tabular}
    \caption{\label{fig:calorimeter} (Left) CAD drawings of the calorimeter disks.
      Calorimeter innermost (outermost) radius is of 350 mm (600 mm).
      Layout of the FEE and digitization crates is also shown. (Right) Exploded view of all the calorimeter parts.}
\end{center}
\end{figure}

In Figure~\ref{fig:calorimeter} (left) a drawing of these two annuli is shown.
Similarly to the tracker, the inner circular hole allows 
electrons up to 55 MeV$/$c momenta to escape undetected. 

In Figure~\ref{fig:calorimeter} (right) an exploded view of a single calorimeter annulus is shown.\\
It consists of an outer monolithyc Al cylinder that provides the main support for the crystals and integrates the feet and adjustment mechanism to park the detector on the rails inside the detector solenoid. The inner support is made of a Carbon Fiber  cylinder that maximise passive material $X_0$.
The crystals are then sandwiched between two cover plates. A  Carbon Fiber front plate  also integrates thin wall Al pipes to flow the radioactive Fluorinert fluid to calibrate the response; a back  plate made of PEEK with apertures in correspondence of each crystals where the Front End Electronics (FEE) and SiPM holders will be inserted.
The back plate houses also the Copper pipes where a coolant is flown to thermalise the photosensors and extract the power dissipated by both the FEE and the sensors.
The calorimeter has to operate in the hostile experimental environment with  1 $T$ magnetic field and a vacuum of $10^{-4}$ Torr, a maximum neutron fluence of $10^{12} \;n/$cm$^2$ in 3 years, a maximum ionising dose of 100 krad in the hottest region at lower radii of the calorimeter.
10 custom made crates are arranged on top of the outer cylinder and are connected to the cooling circuit.

The calorimeter particle identification provides a good separation between CE's 
and muons, un-vetoed by the CRV,  mimicking the signal. 
The required muon rejection factor (200) is achieved with 95\% efficiency on the signal, 
combining the time of flight difference between the tracker track and the calorimeter 
cluster with the E$/$p ratio.

In order to satisfy these requirements, the calorimeter has to reach
an energy resolution of O($5 \%$), a time resolution less than 500~ps and
a position resolution better than 1~cm for 100~MeV electrons. The selected
crystals should also be radiation hard up to 100 krad. The photosensors are shielded
by the crystals themselves and should only sustain a fluency up to
$3 \times 10^{11}$~n/cm$^2$. 

\subsection{Calorimeter performances and prototyping}
\begin{figure}
  \begin{center}
    \begin{tabular}{ccc} \\
      \includegraphics[width=0.3\columnwidth]{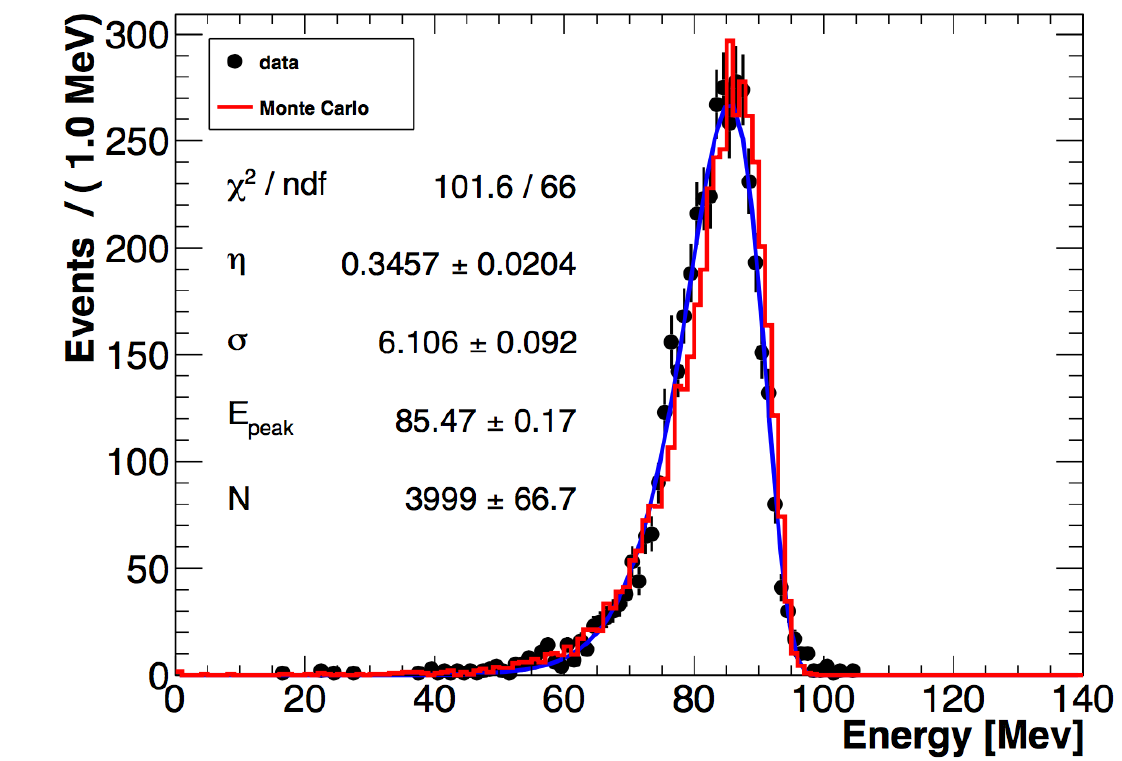}  &
      \includegraphics[width=0.3\columnwidth]{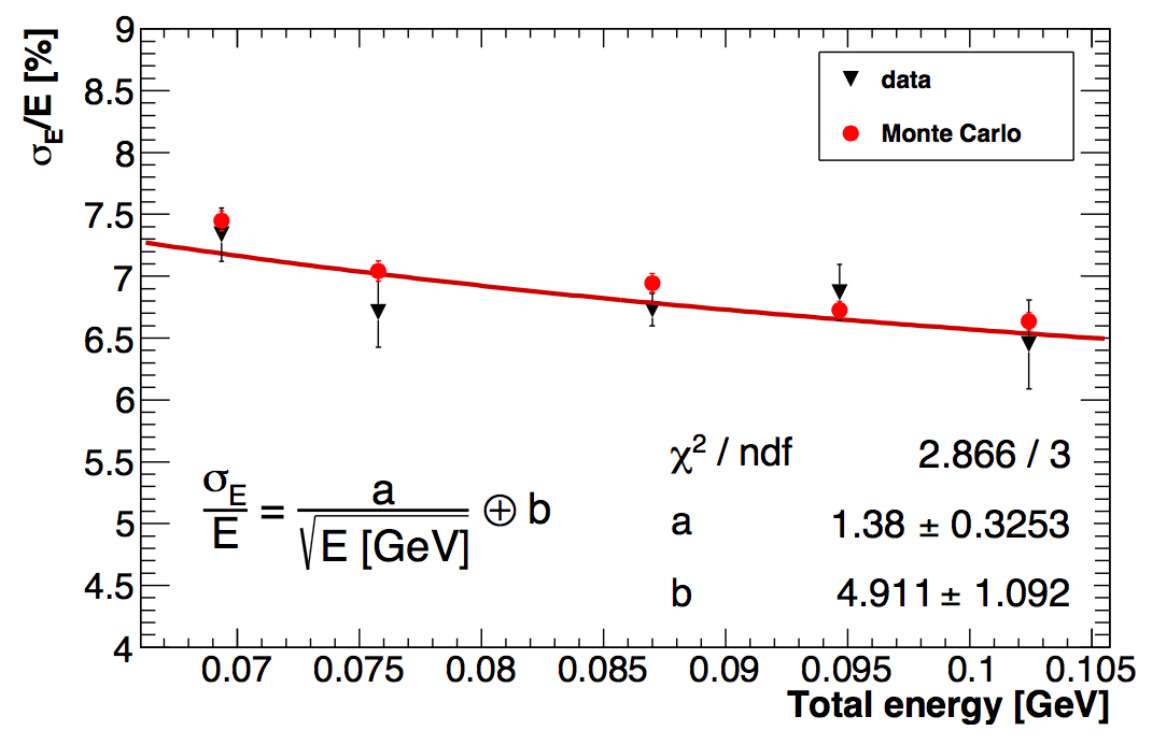}&
        \includegraphics[width=0.3\columnwidth]{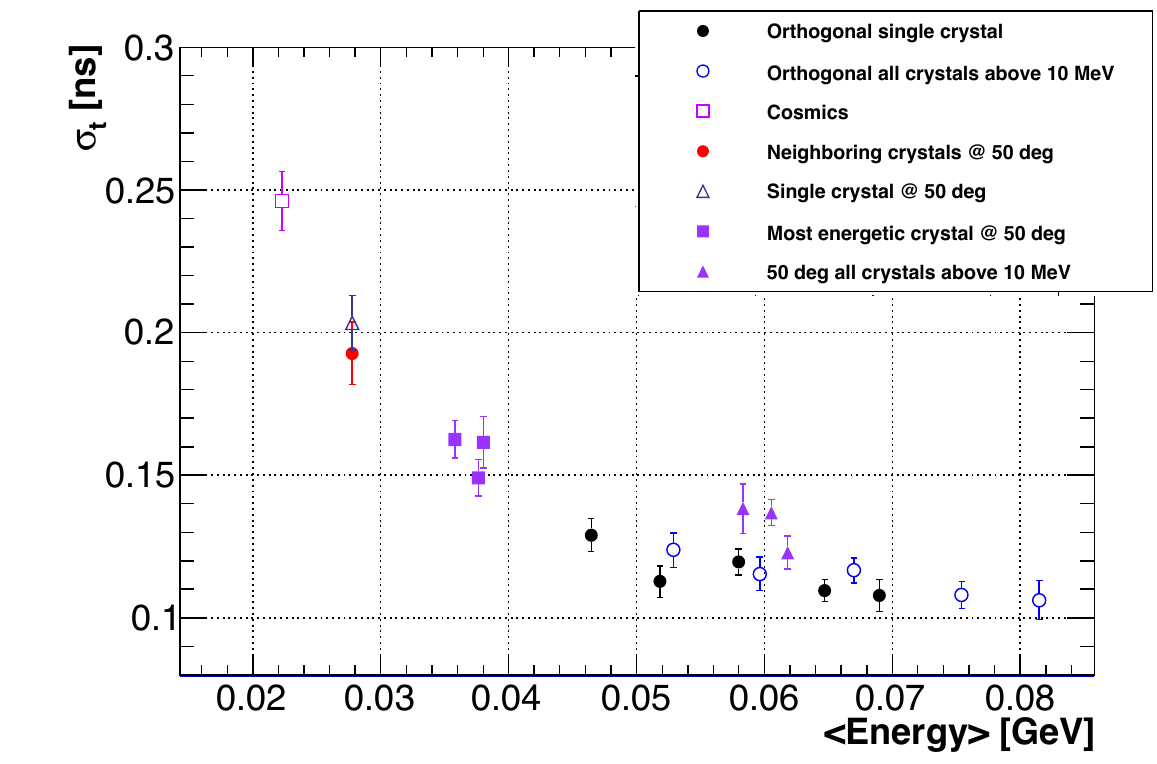}\\
    \end{tabular}
    \caption{\label{fig:btf} BTF results:(left) Data-MC comparison for energy reconstruction of 100 MeV electrons; a typical fit to the data with a log-normal function is shown in red; left long tail is due to not full containment of the shower;
    (center) energy resolution as a function of the reconstructed total energy; (right) time resolution as function of  energy for different configurations of the matrix as function of energy.}
\end{center}
\end{figure}

A calorimeter prototype consisting of a $3 \times 3$ matrix of $30 \times 30 \times 200$ mm$^2$  un-doped CsI crystals wrapped with 150 $\mu$m Tyvek and read out by one $12 \times 12$ mm$^2$ SPL TSV SiPM by Hamamatsu
has been tested with an electron beam at the Beam Test Facility (BTF) in Frascati  during
April 2015. 
The results, described in \cite{btf}, are coherent with the ones predicted by the GEANT4  simulation ~\cite{geant} and are shown in Figure~\ref{fig:btf}:
\begin{itemize}
\item time resolution better than 150 $p$s for 100 MeV electrons. The timing resolution ranges from about 250 ps
at 22 MeV to about 120 ps in the energy range above 50 MeV.
\item energy resolution of $\sim7\% $ for 100 MeV electrons, dominated by the shower non-containment.
\end{itemize}

\begin{figure}
  \begin{center}
    \begin{tabular}{cc} \\
      \includegraphics[width=0.6\columnwidth]{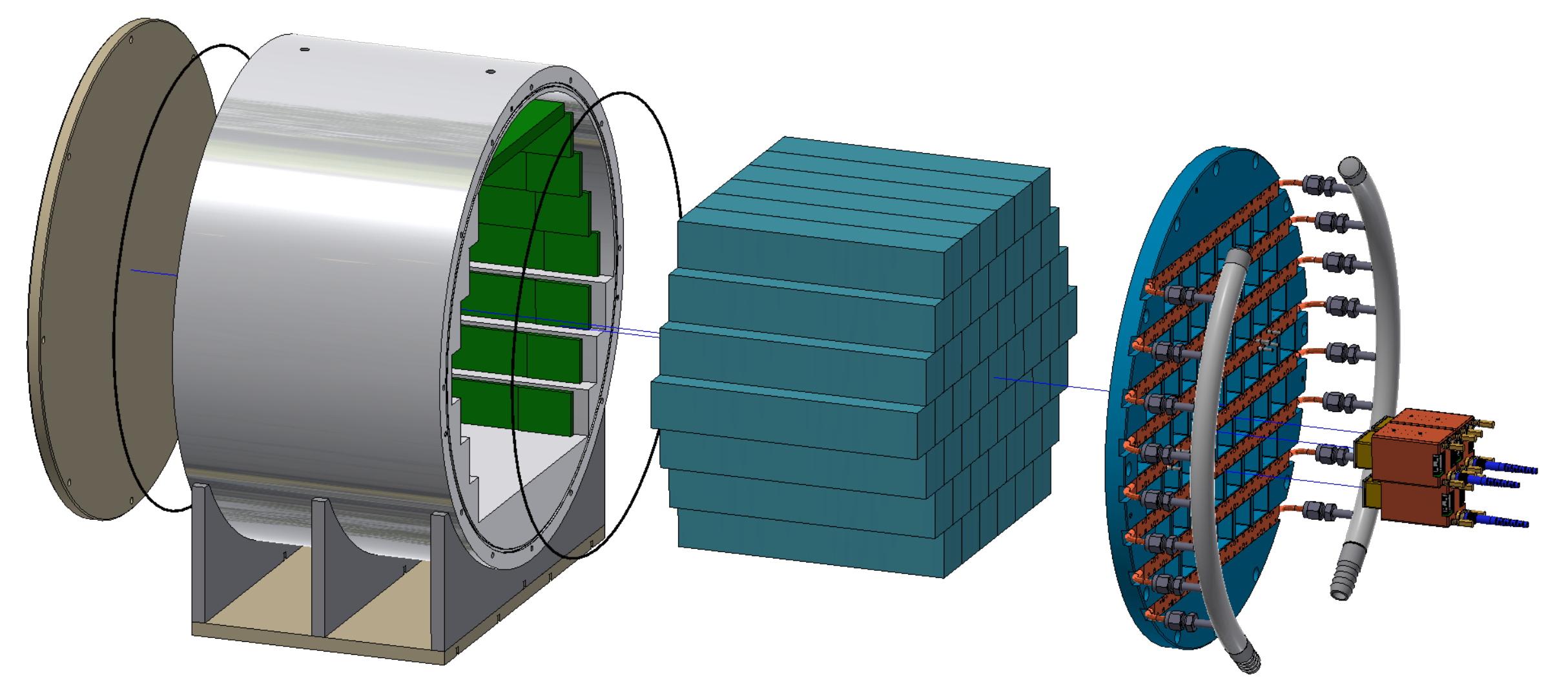}  &
      \includegraphics[width=0.4\columnwidth]{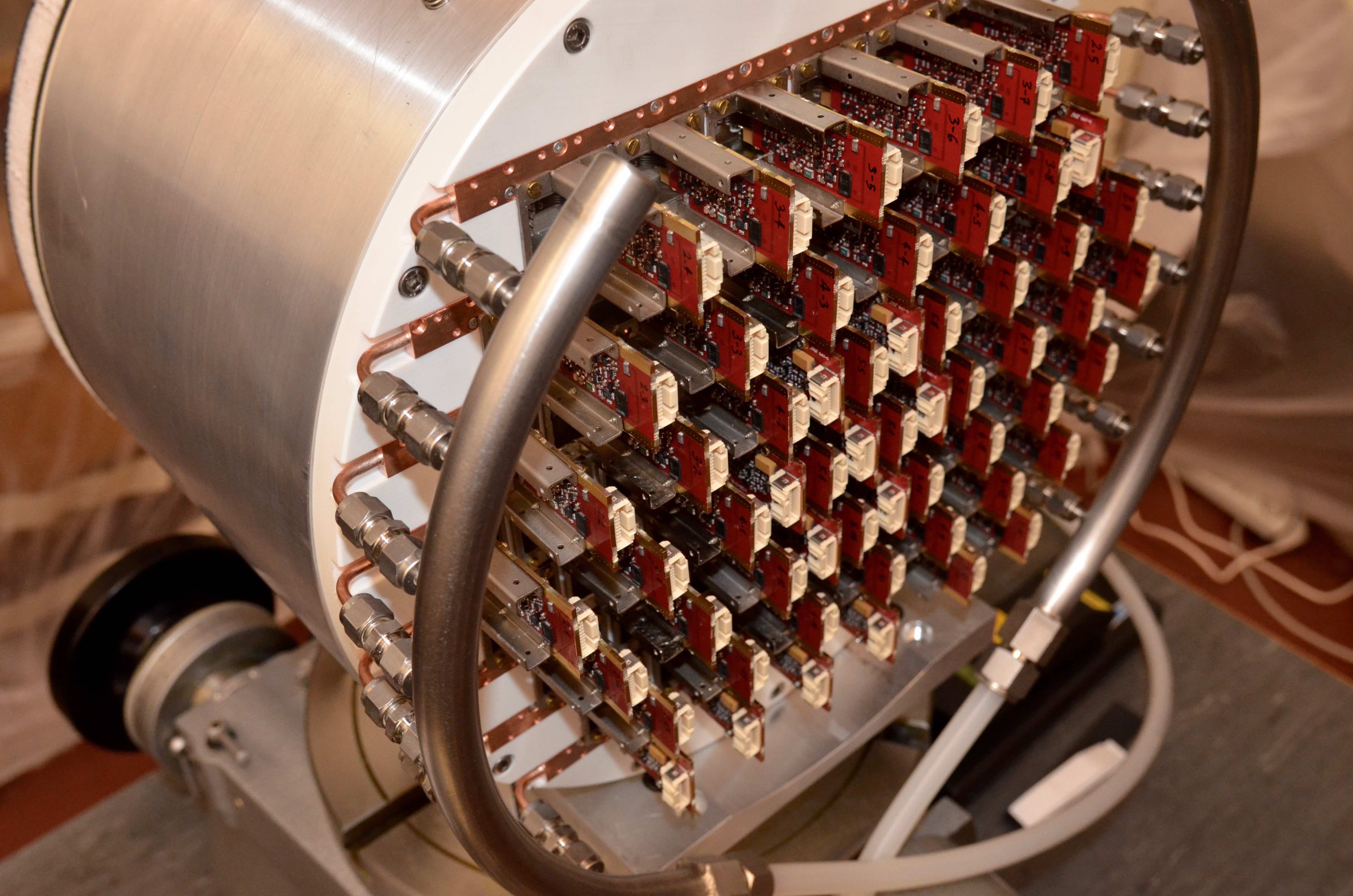} \\
    \end{tabular}
     \caption{\label{fig:module0} CAD drawing of the Module-0 (left) and its realization (right)}
\end{center}
\end{figure}

We have built a Module-0 prototype composed of 51 CsI crystals from different vendors (Siccas, St. Gobain and Amcrys) instrumented with SiPM from 3 different companies (Hamamatsu, Sensl, Advansid) to test their quality and to test the current design technological performance. Figure ~\ref{fig:module0}  shows the CAD drawing of the Module-0 and its actual realization. This Module-0 has been tested at the Frascati BTF and data analysis is underway.

\begin{figure}
  \begin{center}
    \begin{tabular}{ccc} \\
      \includegraphics[width=0.3\columnwidth]{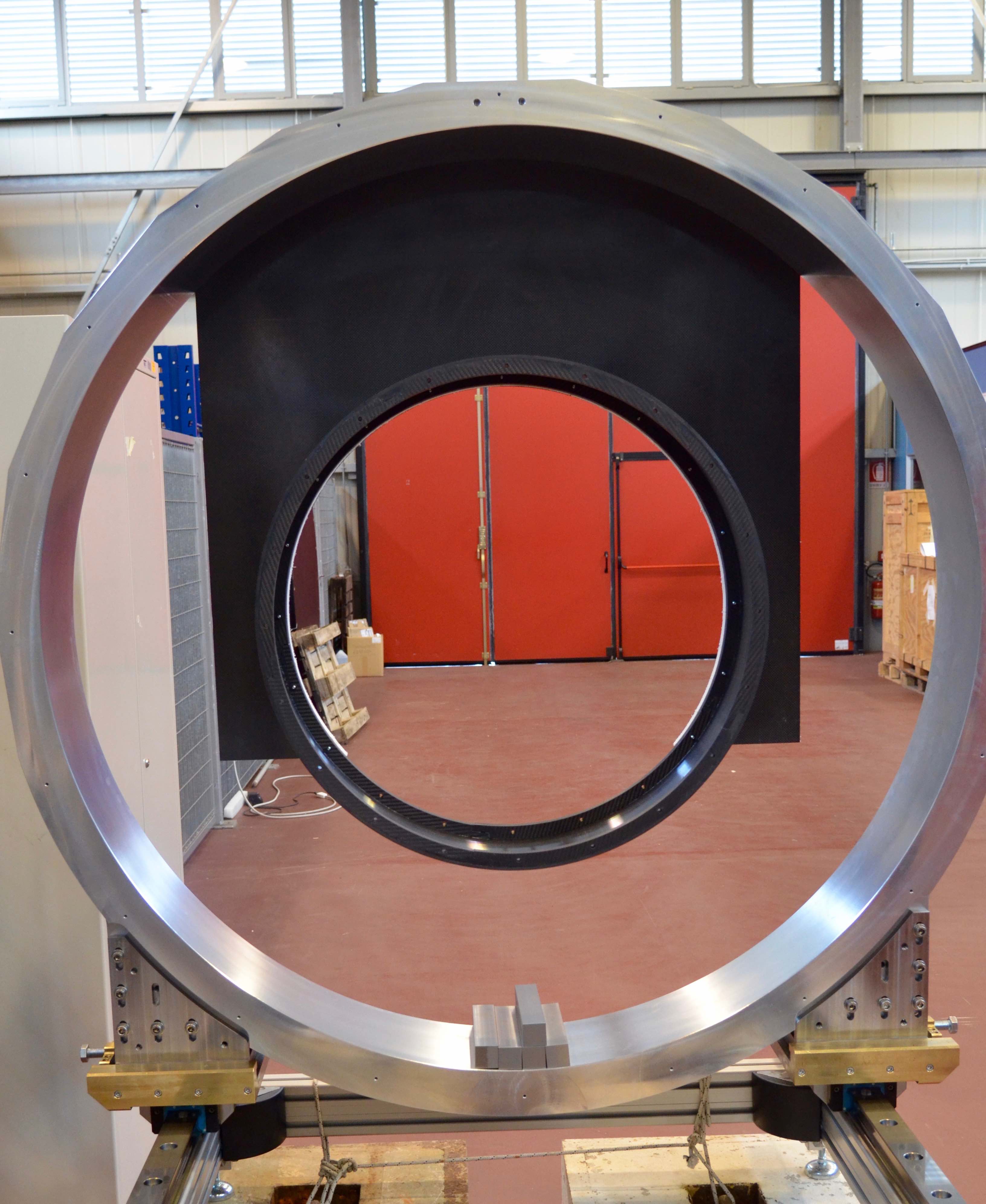}  &
      \includegraphics[width=0.3\columnwidth]{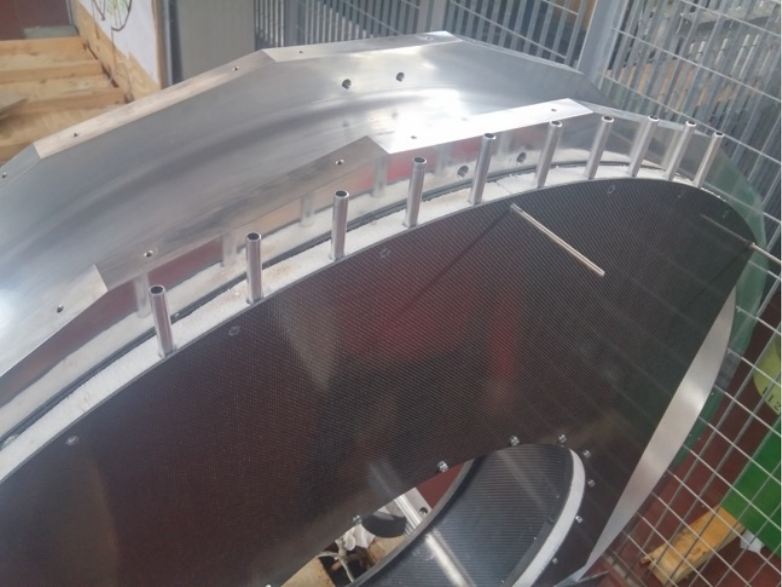} &
      \includegraphics[width=0.3\columnwidth]{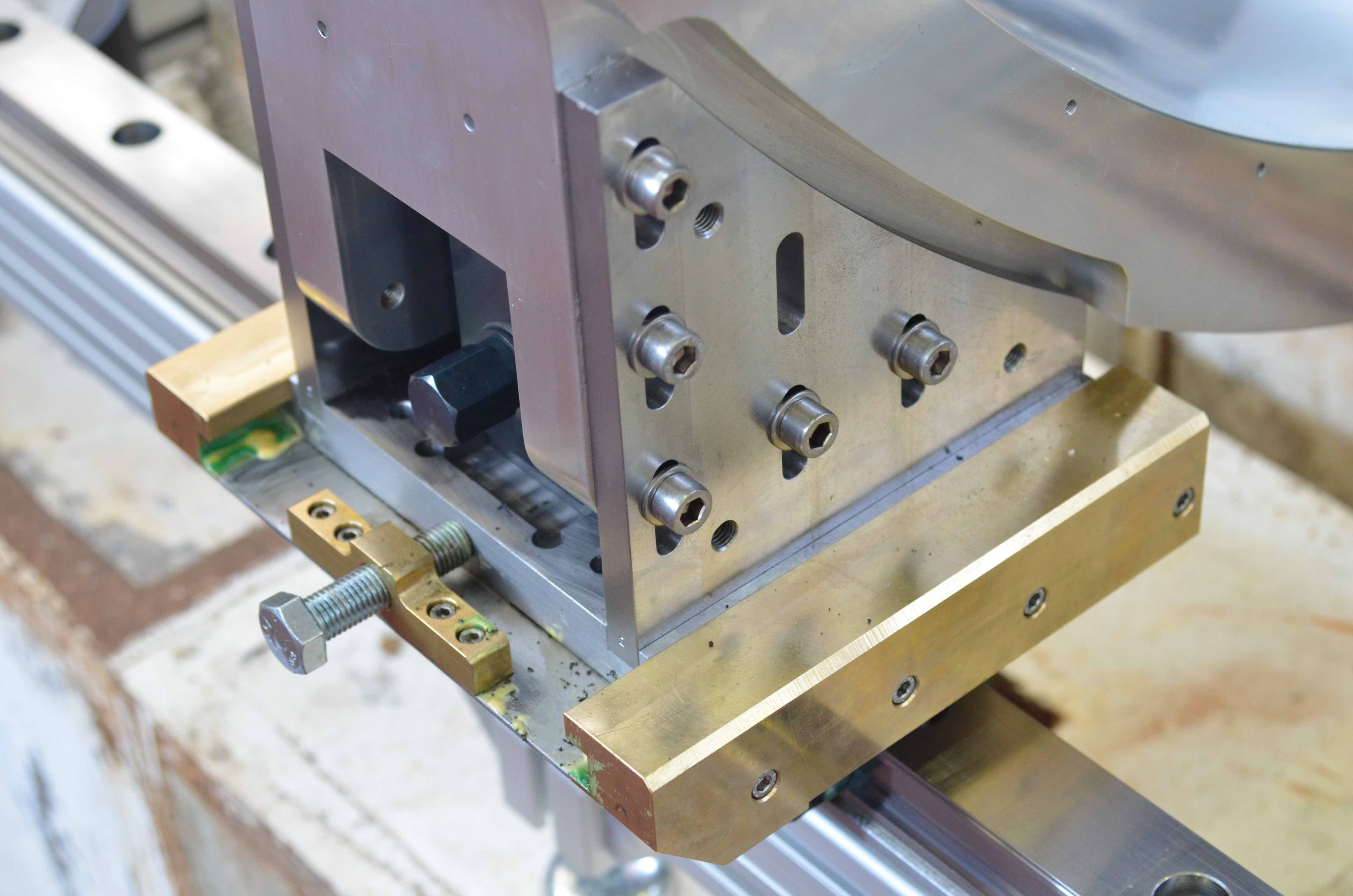} \\
    \end{tabular}
     \caption{\label{fig:mockup} Full scale mock-up. Outer Al cylinder, front plate with source piping, foot with x-y adjustment.}
\end{center}
\end{figure}

A full scale mock-up of the mechanical structure is being built to test the assembly of the crystals, FEE electronics, cooling system and overall structure robustness : the Al outer ring, the inner Carbon Fiber cylinder, sections of the front and back plates, crate prototype have been manufactured.
A whole annulus will be assembled using a mixture of fake Iron crystals and a sample of preproduction CsI crystals. Figure~\ref{fig:mockup} shows the ongoing mock-up.

\section{Conclusions and perspectives}
The Mu2e experiment design and construction proceeds well and
it is on schedule to be commissioned with beam for the end of 2021.
Its goal is to probe CLFV with a single event sensitivity of 2.5 $\times$
10$^{-16}$ or set an upper limit on the conversion rate  $< 6 \times 10^{-17}$
at 90 \% C.L. improving of four orders of magnitude the 
sensitivity of previuos measurements . 
A Mu2e second phase is already planned with the goal of increasing the sensitivity
of an additional factor of 10.    
The Calorimeter design is almost complete and the prototyping of the most delicate components is underway. The Module-0 construction showed a good coupling between photosensors and crystals and also the capability to cool down the SiPM and extract heat with the current cooling scheme.
We will start the construction of the final components in Year 2018 together with the opening of the tenders for crystals and SiPM purchase.

\acknowledgments
This work was supported by the EU Horizon 2020 Research and Innovation Programme under the Marie Sklodowska-Curie Grant Agreement No. 690835.


\end{document}